\documentclass[aps,pra,twocolumn,floatfix,footinbib,notitlepage,superscriptaddress,groupaddress,showpacs]{revtex4-1}
\usepackage{times,graphics,graphicx,xcolor,amsfonts,amssymb,amsbsy,amsthm,amsmath,hyperref,bm,bbm,color,makecell,dsfont,upgreek}
\hypersetup{colorlinks,linkcolor={blue},citecolor={blue},urlcolor= {blue}}
\newcommand{\ket}[1]{\left|{#1}\right\rangle}

\newcommand{\ketbra}[2]{|{#1}\rangle \langle{#2}|}
\newcommand{\eeqref}[1]{Eq.~(\ref{#1})}
\newcommand{\figref}[1]{Fig.~\ref{#1}}

\newcommand{\ignore}[1]{}

\let\oldsqrt\sqrt
\def\sqrt{\mathpalette\DHLhksqrt}
\def\DHLhksqrt#1#2{%
\setbox0=\hbox{$#1\oldsqrt{#2\,}$}\dimen0=\ht0
\advance\dimen0-0.2\ht0
\setbox2=\hbox{\vrule height\ht0 depth -\dimen0}%
{\box0\lower0.4pt\box2}}

\DeclareFontFamily{OT1}{pzc}{}
\DeclareFontShape{OT1}{pzc}{m}{it}%
              {<-> s * [1.25] pzcmi7t}{}
\DeclareMathAlphabet{\mathpzc}{OT1}{pzc}%
                                 {m}{it}

\begin{document}

\title{Quantum imaging with undetected photons as ancilla-assisted process tomography}

\author{M. Ghalaii}
\affiliation{Department of Physics, Sharif University of Technology, Tehran 14588, Iran}
\affiliation{School of Electronic and Electrical Engineering, University of Leeds, Leeds LS2 9JT, UK}

\author{M. Afsary}
\affiliation{Department of Physics, Sharif University of Technology, Tehran 14588, Iran}

\author{S. Alipour}
\affiliation{School of Nano Science, Institute for Research in Fundamental Sciences (IPM), Tehran 19538, Iran}
\affiliation{Department of Physics, Sharif University of Technology, Tehran 14588, Iran}

\author{A. T. Rezakhani}
\email{rezakhani@sharif.edu}
\affiliation{Department of Physics, Sharif University of Technology, Tehran 14588, Iran}
\affiliation{School of Nano Science, Institute for Research in Fundamental Sciences (IPM), Tehran 19538, Iran}

\begin{abstract}
We show that how a recent experiment of quantum imaging with undetected photons can basically be described as a (partial) ancilla-assisted process tomography. We propose a simplified quantum circuit version of this scenario, which also enables to recast quantum imaging in quantum computation language. Our analogy and analysis may help better understand the role of classical and/or quantum correlations in imaging experiments.
\end{abstract}

\pacs{03.67.-a, 03.65.Wj, 42.30.Wb, 42.50.Ex}
\date{\today}
\maketitle

\section{Introduction}
\label{sec:int}

Quantum imaging (QI) is an interesting technique which employs quantum properties of light, such as entanglement and nonlocality, to give a high resolution image from a partially-transmitting object \cite{Klyshko-1}. QI has also been experimentally demonstrated in numerous experiments \cite{Pittman}, e.g., through the method of ``ghost imaging" \cite{Gatti-1,Shih,Shapiro,Aspden}.

To recover information about an unknown `object' (illuminated by either a ``signal" or an ``idler" photon), most QI methods employ entanglement of quantum states generated by parametric down-conversion. However, it has been shown that entanglement is not necessary for ghost imaging since using classically-correlated fields \cite{Bennink}, thermal  \cite{Gatti-2}, and pseudo-thermal lights \cite{Ferri,Valencia} still enables obtaining images through ghost imaging. In addition, it has been shown that using half-wave plates properly (considering polarization as a degree of freedom), one can affect output probabilities and thus visibility of the images \cite{which-way}. Recently, a novel and elegant method for QI has been experimentally demonstrated in Ref.~\cite{Lemos}, where ``which-path entanglement" seems to be crucial in order to have image. A closer theoretical inspection of this experiment from the perspective of quantum optics has been reported in Ref.~\cite{Lahiri}.

Quantum process tomography (QPT) is another technique employed to identify unknown quantum ``processes" \cite{Nielsen}. Several schemes have been outlined to accomplish this task, namely, standard QPT \cite{Nielsen,Chuang,Poyatos}, ancilla-assisted process tomography (AAPT) \cite{Leung,D'Ariano-AAPT-1,Altepeter,D'Ariano-AAPT-2}, and direct characterization of quantum dynamics \cite{Mohseni-DCQD,DCQD-corr,Wang-DCQD}---see Ref.~\cite{Mohseni} for an extensive review. Notwithstanding their differences, all QPT methods operate similarly based on probing an unknown process (as a ``black box") with appropriate input states, measuring output states, and identifying the process through relation of input and output states. 

Establishing a connection between quantum tomography and \textit{imaging} or \textit{sensing} schemes can be interesting from various aspects. It has been argued that tomography and spectroscopy can be considered as dual forms of quantum computation \cite{Miquel-etal-Nature}. Following a similar reasoning, one may also argue that QPT and QI of an object are basically akin in the sense that they both examine a(n) process/object with probe states and then analyze output states. Here we make this connection more explicit. In particular, we demonstrate that the QI scheme proposed in Ref.~\cite{Lemos} can be described as a version of AAPT in which the object is assumed in both as a black box, whereas an identical or partial ``image" is obtained by analyzing ancillary probes. We elaborate in detail how this analogy work through identifying different steps of the QI scheme with preparation and measurement parts of a special AAPT scenario. Additionally, we represent a quantum circuit for the QI scheme, which helps analyze the role of initial-state correlations in QI. We demonstrate that, by replacing the entangled Bell states with arbitrary Werner states, QI can also work by non-entangled light. 

This manuscript is structured as follows. In Sec.~\ref{sec:the-imaging-setup}, we review the QI scheme of Ref.~\cite{Lemos}. In Sec.~\ref{sec:QI-AAPT}, we suggest a quantum circuit version of the QI scenario in which only one-qubit and \textsc{cnot} gates as well as a measurement are used. In Sec.~\ref{sec:AAPT-UP}, we describe an analogy between the QI scheme and AAPT by translating steps of the QI scheme into a version of AAPT in which parts of the probe systems remain undetected. We explicitly show that which parts of the QI scheme correspond to preparation and measurement parts of the AAPT scheme. In addition, we show that quantum entanglement of probe states is not essential in generating an image, and that how one can supplement measurements to enable a full tomography of the object. We summarize our findings in Sec.~\ref{sec:conc}. There are several appendices which include details of parts of calculations.

\section{The imaging setup}
\label{sec:the-imaging-setup}

Here we focus on a recent method and experiment for creating ``image" of an object in a quantum mechanical fashion, proposed in Ref.~\cite{Lemos} and theoretically furthered in Ref.~\cite{Lahiri}). Our discussion here concerns (a simplified version of) this experiment without realization technicalities, and aims at explaining how imaging an ``object" can be understood in the framework of QPT.

\begin{figure}[tp]
\includegraphics[scale=0.36]{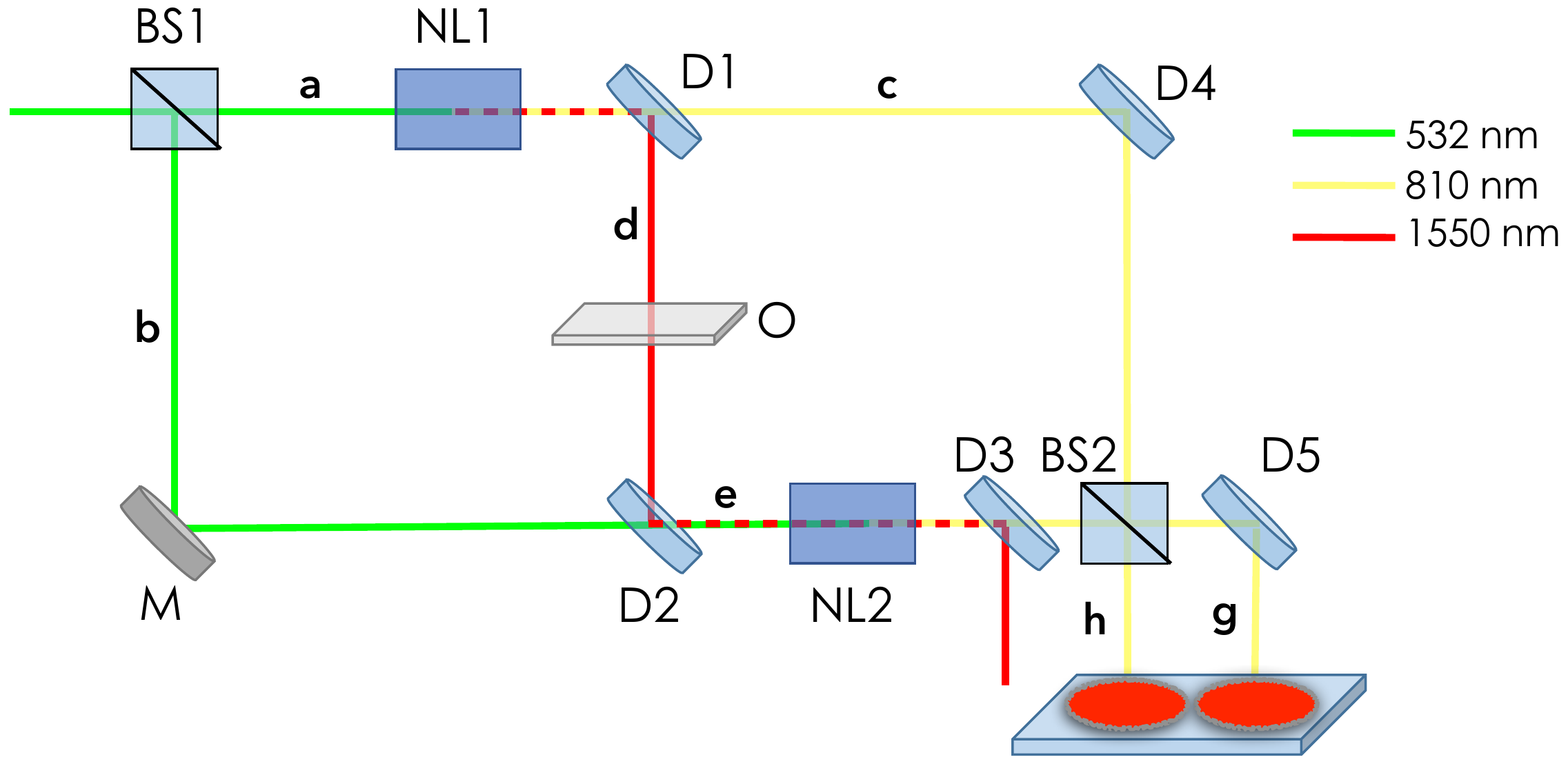}
\caption{(color online). Experimental setup of the QI scenario in Ref.~\cite{Lemos}.}
\label{fig:setup}
\label{setup}
\end{figure}

In this experimental setup a beam splitter (BS1), which is illuminated by a \textit{pump} photon, is used to generate the following path-entangled Bell state:
\begin{equation}
\label{BS1-action}
\ket{\Psi_1}=\big(|1\rangle_{p,\mathbf{a}}|0\rangle_{p,\mathbf{b}} +|0\rangle_{p,\mathbf{a}}|1\rangle_{p,\mathbf{b}}\big)/\sqrt{2},
\end{equation}
where $p$ indicates the wavelength of the pump photon, and $\mathbf{a}$ and $\mathbf{b}$ denote different paths which the photon can choose (Fig.~\ref{setup}). To set our notation hereafter, we assume that $\ket{n}_{\lambda,\mathbf{t}}$ indicates an $n$-photon state with wavelength $\lambda$ in path $\mathbf{t}$. Since at any instant there exists at most one photon in each \textit{mode}, we use the encoding $\ket{n=0}\equiv (1~~0)^{T}$ and $\ket{n=1}\equiv (0~~1)^{T}$, which represent a logical basis as a ``qubit" \cite{Nielsen}.

This setup includes two nonlinear crystals used to generate extra photons. These crystals should be pumped identically but not simultaneously. While passing through the nonlinear medium (NL) \cite{bib:NL}, a pump photon ($532$ nm, in the original experiment) can convert to a \textit{signal} ($s$) and an \textit{idler} ($i$) photon with different wavelengths ($810$ nm and $1550$ nm, respectively, in the original experiment) through a down-conversion process. This is a probabilistic event. Hence the probability of obtaining a photon in the final detectors of the setup are conditioned on a successful down-conversion process. However, in the following, we do not consider unsuccessful down-conversion events as they do not yield an image in this setup. NL1 changes the Bell state (\ref{BS1-action}) to
\begin{align}
\ket{\Psi_2}=\frac{1}{\sqrt{2}}(|1\rangle_{s,\mathbf{a}} |1\rangle_{i,\mathbf{a}}|0\rangle_{p,\mathbf{a}} |0\rangle_{p,\mathbf{b}}+ |0\rangle_{s,\mathbf{a}}|0\rangle_{i,\mathbf{a}}|0\rangle_{p,\mathbf{a}} |1\rangle_{p,\mathbf{b}}).
\end{align}
The signal and idler photons are next separated (because of their distinct wavelengths) by a dichroic mirror (D1), which reflects the idler photons into path $\mathbf{d}$ and allows photons with other wavelengths to pass. This yields
\begin{align}
\label{bef-ob}
\ket{\Psi_3}=\frac{1}{\sqrt{2}}(|1\rangle_{s,\mathbf{c}} |1\rangle_{i,\mathbf{d}}|0\rangle_{p,\mathbf{c}} |0\rangle_{p,\mathbf{b}}
+|0\rangle_{s,\mathbf{c}}|0\rangle_{i,\mathbf{d}}|0\rangle_{p,\mathbf{c}} |1\rangle_{p,\mathbf{b}}).
\end{align}
The idler photon then illuminates the object located in path $\mathbf{d}$; it either passes through the object ($O$) with the associated transmission factor $Te^{i\gamma}$, or is reflected with the associated reflectivity factor $\sqrt{1-T^2}$. 

We note that for a real two-dimensional object both $T$ and $\gamma$ depend on position on the object, that is, $T=T(x,y)$ and $\gamma=\gamma(x,y)$, where $(x,y)\in O$. In the real experimental setup of Ref.~\cite{Lemos}, the object is placed between two lenses, and it has been argued that \cite{Lahiri} an incident plane wave mode with wave vector $\overrightarrow{\mathbf{k}}$ can illuminate one point of the object and hence yields an output wave with an associated wave vector $\overrightarrow{\mathbf{k}'}(\overrightarrow{\mathbf{k}})$. Given that a quantized light is a superposition of plane wave modes, one can assume that one point on the object can transmit and reflect only one specific mode of the quantized (idler) beam \cite{Lahiri}. Bearing in mind this one-to-one correspondence between the object and the image, it thus suffices---for our purposes here---to explain the basic theory of the imaging scenario through a single point of the object with constant $T$ and $\gamma$ factors. One may consider this restricted sort of imaging as a ``sensing" scenario.

After interacting with the object, $|\Psi_3\rangle$ changes to
\begin{align}
\ket{\Psi_4}=\frac{1}{\sqrt{2}}\big(&Te^{i\gamma}|1\rangle_{s,\mathbf{c}} |1\rangle_{i,\mathbf{d}}|0\rangle_{i,\mathbf{w}}|0\rangle_{p,\mathbf{b}}|0\rangle_{p,\mathbf{c}}\nonumber\\
&+\sqrt{1-T^2}|1\rangle_{s,\mathbf{c}}|0\rangle_{i,\mathbf{d}} |1\rangle_{i,\mathbf{w}}|0\rangle_{p,\mathbf{b}}|0\rangle_{p,\mathbf{c}}\nonumber \\
&+|0\rangle_{s,\mathbf{c}}|0\rangle_{i,\mathbf{d}}|0\rangle_{i,\mathbf{w}}|1\rangle_{p,\mathbf{b}}|0\rangle_{p,\mathbf{c}}\big),
\end{align}
where $\mathbf{w}$ is the path through which the reflected idler photon passes. There is a probability that after BS1 the photon goes through path $\mathbf{b}$ and the whole state be affected by the dichroic mirror D2. This mirror maps paths $\mathbf{d},\mathbf{b}\to \mathbf{e}$, thus the state of the system is transformed to 
\begin{align}
&\frac{1}{\sqrt{2}}\Big(Te^{i\gamma}|1\rangle_{s,\mathbf{c}} |1\rangle_{i,\mathbf{e}} |0\rangle_{i,\mathbf{w}} |0\rangle_{p,\mathbf{e}} |0\rangle_{p,\mathbf{c}} \nonumber\\
&~+ \sqrt{1-T^2} |1\rangle_{s,\mathbf{c}} |0\rangle_{i,\mathbf{e}} |1\rangle_{i,\mathbf{w}} |0\rangle_{p,\mathbf{e}} |0\rangle_{p,\mathbf{c}}\nonumber \\
&~+|0\rangle_{s,\mathbf{c}} |0\rangle_{i,\mathbf{e}} |0\rangle_{i,\mathbf{w}} |1\rangle_{p,\mathbf{e}} |0\rangle_{p,\mathbf{c}}\Big).
\end{align} 
Now NL2 acts on this state as $|0\rangle_{s,\mathbf{e}} |0\rangle_{i,\mathbf{e}} |1\rangle_{p,\mathbf{e}}\to |1\rangle_{s,\mathbf{e}} |1\rangle_{i,\mathbf{e}} |0\rangle_{p,\mathbf{e}}$; that is, it transforms a photon with the pump frequency to two photons along the same path $\mathbf{e}$. Hence the total state becomes
\begin{align}
\ket{\Psi_5}=&\frac{1}{\sqrt{2}}\Big(Te^{i\gamma}|1\rangle_{s,\mathbf{c}} |0\rangle_{s,\mathbf{e}} |1\rangle_{i,\mathbf{e}} |0\rangle_{i,\mathbf{w}} |0\rangle_{p,\mathbf{e}} |0\rangle_{p,\mathbf{c}}\nonumber\\
&~+\sqrt{1-T^2} |1\rangle_{s,\mathbf{c}} |0\rangle_{s,\mathbf{e}} |0\rangle_{i,\mathbf{e}} |1\rangle_{i,\mathbf{w}} |0\rangle_{p,\mathbf{e}} |0\rangle_{p,\mathbf{c}}\nonumber \\
&~+|0\rangle_{s,\mathbf{c}} |1\rangle_{s,\mathbf{e}} |1\rangle_{i,\mathbf{e}} |0\rangle_{i,\mathbf{w}} |0\rangle_{p,\mathbf{e}} |0\rangle_{p,\mathbf{c}}\Big).
\end{align}
By using another dichroic mirror (D3), the idler photons are discarded from the setup. The pump photons are also discarded through D4 and D5; the reflected photon from the object is discarded too. Mathematically, such discarding is represented by tracing out the idler photon, the pump photon, and the photon reflected to path $\mathbf{w}$. Hence, the total state before BS2 becomes
\begin{align}
\label{bef-BS2}
{\varrho}_6 =&\frac{1}{2}\big(|1\rangle_{\mathbf{c}}\langle 1|\otimes
|0\rangle_{\mathbf{e}}\langle 0|+Te^{i\gamma}|1\rangle_{\mathbf{c}}\langle 0|\otimes |0\rangle_{\mathbf{e}}\langle 1|\nonumber   \\
&+Te^{-i\gamma}|0\rangle_{\mathbf{c}}\langle 1|\otimes |1\rangle_{\mathbf{e}}\langle 0|+|0\rangle_{\mathbf{c}}\langle 0|\otimes |1\rangle_{\mathbf{e}}\langle 1|\big),
\end{align}
where to ease the notation we have removed the wavelength indices because after tracing out the idler and pump photons only signal photons remain.

\begin{figure}[bp]
\centering
\includegraphics[scale=.3]{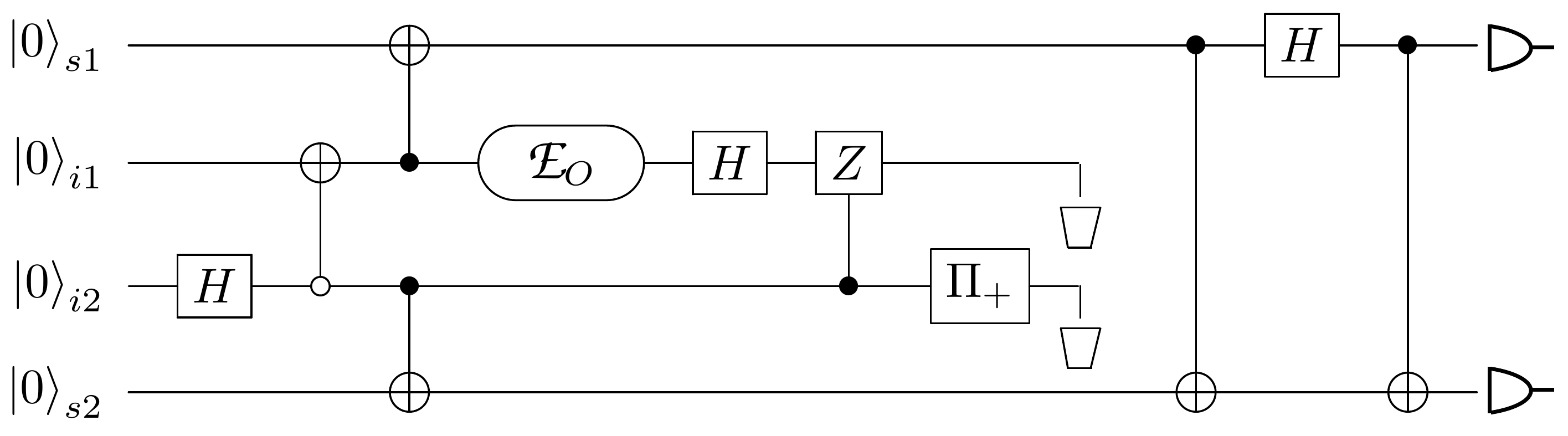}
\caption{Quantum logical circuit of the imaging setup of Fig.~\ref{fig:setup}. Here $H$ is the Hadamard gate and $Z$ is the $z$-Pauli gate.}
\label{logic-circuit}
\end{figure}

A pivotal feature of this setup is that after NL2 the source of the idler photons cannot be distinguished, and this fundamental ambiguity in ``which-path information" is indeed responsible for creating the image. In other words, $|1\rangle_{i,\mathbf{d}}\equiv|1\rangle_{i,\mathbf{e}}$, which indicates that after NL2 the source as well as the path of the idler photons cannot be discerned. At the final stage, BS2 is applied to combine the information carried in paths $\mathbf{c}$ and $\mathbf{e}$, and to generate the image. The probability of finding a photon in paths $\mathbf{h}$ and $\mathbf{g}$ is obtained as
\begin{equation}
\label{prob-main-setup}
P_{\mathbf{h}/\mathbf{g}}=(1\mp T\cos\gamma )/2.
\end{equation}

\section{The imaging setup represented by a quantum circuit}
\label{sec:QI-AAPT}

In this section, we present a quantum circuit which can simulate the setup of Fig.~\ref{fig:setup}. 

We assume a quantum circuit as in Fig.~\ref{logic-circuit}, which is applied on four \textit{qubits} prepared in the different field modes $\ket{\Phi_0}=|0\rangle_{s1}|0\rangle_{i1}|0\rangle_{i2}|0\rangle_{s2}$, where indices ``$1$" and ``$2$" imply the source of the qubits (or photons), namely,  NL1 and NL2. The first Hadamard and the three subsequent \textsc{cnot} gates turn $\ket{\Phi_0}$ to (App. \ref{psi_0-psi_prob})
\begin{equation}
\label{Psi-probe}
\ket{\Phi_{\mathrm{probe}}}=\big(|1\rangle_{s1}|1\rangle_{i1}|0\rangle_{i2}|0\rangle_{s2} + |0\rangle_{s1}|0\rangle_{i1} |1\rangle_{i2}|1\rangle_{s2} \big)/\sqrt{2}.
\end{equation}
The Hadamard followed by the first \textsc{cnot} (which acts when the control qubit is $|0\rangle$) play the role of BS1, and the other two \textsc{cnot}s simulate the action of NL1 and NL2 in the experimental setup.

The object ($O$) is represented by the quantum channel (process or operation) $\mathpzc{E}_O$ which acts as (App. \ref{obj-map})
\begin{align}
\label{map}
\mathpzc{E}_O:~\begin{cases}
\ketbra{0}{0}\to \ketbra{0}{0}  \\
\ketbra{0}{1}\to  Te^{-i\gamma}\ketbra{0}{1}  \\
\ketbra{1}{0} \to  Te^{i\gamma}\ketbra{1}{0}  \\
\ketbra{1}{1} \to T^2 \ketbra{1}{1}+ (1-T^2)\ketbra{0}{0}.
\end{cases}
\end{align}
After applying $\mathpzc{E}_O$, we need some operation which can make $|1\rangle_{i,\mathbf{d}} \equiv|1\rangle_{i,\mathbf{e}}$ in the original experimental setup. In our circuit, this can be achieved, for example, if we make the two states $|1\rangle_{i_{1}}|0\rangle_{i_{2}}$ and $|0\rangle_{i_{1}}|1\rangle_{i_{2}}$ quantum-mechanically indistinguishable. To do so, we can consider a \textit{mode mixer} (MM) such that
\begin{equation}
\mathrm{MM}:~~\begin{cases}|0\rangle_{i_1} |1\rangle_{i_2}\\ |1\rangle_{i_1} |0\rangle_{i_2} \end{cases}\to |\Xi\rangle_{i_{1} i_{2}},
\label{MM-def}
\end{equation}
and acts as identity otherwise, where $|\Xi\rangle$ is a fixed state of the idlers (App. \ref{object-app}). The MM can be realized in various manners. As an example, consider one Hadamard gate, one \textsc{cz} gate, and the projection $\Pi_+=\ketbra{+}{+}$, applied respectively, where $|\pm\rangle=(1/\sqrt{2})(|0\rangle \pm |1\rangle)$. For this specific operation we have $|\Xi\rangle_{i_{1}i_{2}}=|-\rangle_{i_1}|+\rangle_{i_2}$. In general, the action of MM can be described through the following quantum operation: 
\begin{equation}
\mathpzc{F}_{\mathrm{MM}}[\varrho] = \frac{(\mathsf{P}_{\Xi}+\mathsf{Q})\varrho (\mathsf{P}_{\Xi}+\mathsf{Q})^{\dag}}{\mathrm{Tr}[(\mathsf{P}_{\Xi}+\mathsf{Q})\varrho (\mathsf{P}_{\Xi}+\mathsf{Q})^{\dag}]},
\end{equation}
where $\mathsf{P}_{\Xi}=|\Xi\rangle(\langle 01| + \langle 10|)$ and $\mathsf{Q}= (|00\rangle + |11\rangle)(\langle 00|+ \langle 11|)$. 

The action of MM and tracing out over the idler photons (simulating their absorption or loss in Fig.~\ref{fig:setup}) reduce $\ket{\Phi_{\mathrm{probe}}}$ to 
\begin{align}
\label{after-M1}
\Upsilon=& \frac{1}{2}\big(|1\rangle_{s1}\langle 1| \otimes |0\rangle_{s2}\langle 0|   + Te^{i\gamma} |1\rangle_{s1}\langle 0|  \otimes |0\rangle_{s2}\langle 1|  \nonumber  \\
&+ Te^{-i\gamma}|0\rangle_{s1}\langle 1|  \otimes |1\rangle_{s2}\langle 0|  +
|0\rangle_{s1}\langle 0|  \otimes |1\rangle_{s2}\langle 1| \big),
\end{align}
which is equivalent to the state $\varrho_{6}$ [Eq.~(\ref{bef-BS2})] in the original experimental setup. The details can be found in App. \ref{object-app}.

The action of BS2 can be simulated by a Hadamard and two \textsc{cnot} gates on $\Upsilon$. To detect photons in the first and fourth paths with correct probabilities $P_{s_{1}/s_{2}}=(1\mp T \cos\gamma)/2$ [Eq.~(\ref{prob-main-setup})], we shall need an appropriate measurement $\mathpzc{M}$, which we discuss in the next section.

\section{QI vs. AAPT}
\label{sec:AAPT-UP}

Here we revisit the QI scenario and demonstrate that it can be captured as a form of AAPT. We first explain a general framework for AAPT with \textit{undetected} particles. Next we translate the QI setup into this modified AAPT language. Here by ``undetected" we mean that we discard part of the whole system and employ measurement results on the remained parts to infer the unknown channel. 

\subsection{AAPT with undetected particles}
\label{subsec:AAPT-UP-1}

In the \textit{standard} picture of an AAPT scheme \cite{Leung,D'Ariano-AAPT-1,Altepeter,D'Ariano-AAPT-2}, an unknown quantum process (or black box) acts on a specific system which is correlated with an ancillary system. Identification of the process comes through analyzing results of measurements on the whole composite systems \cite{Mohseni}. 

Another variant of AAPT may include undetected system(s), where measurements are instead performed only on ancilla(s)---whereas the rest of the systems are discarded. In such indirect schemes, although measurements are performed on systems that never directly met (i.e., passed through) the black box, because of correlations between the system and ancilla(s), information of the black box would still leak to the ancillas such that one can identify the black box by measuring the ancilla(s). In the following we briefly describe this modified version of AAPT.

An arbitrary state of a bipartite system (\textit{system} + \textit{ancilla}), with the Hilbert space $\mathpzc{H}_{1}\otimes \mathpzc{H}_{2}$, (as input of AAPT) can be written as (Schmidt decomposition)
\begin{align}
{\varrho}^{(\mathrm{in})}=\sum_{\ell=1}^{D^2} {r}_{\ell} {A}_{\ell}\otimes {B}_{\ell},
\end{align}
where ${r}_{\ell}$s are some (nonnegative) numbers, and the $\{{A}_{\ell}\}_{\ell=1}^{D^2}$ ($\{{B}_{\ell}\}_{\ell=1}^{D^2}$) form an orthonormal operator basis on $\mathpzc{H}_1$ ($\mathpzc{H}_2$), satisfying the orthonormalization condition $\mathrm{Tr}[{A}_{\ell} {A}^{\dag}_{\ell'}]=\delta_{\ell \ell'}$ ($\mathrm{Tr}[{B}_{\ell} {B}^{\dag}_{\ell'}]=\delta_{\ell \ell'}$), with $D$ being the dimension of the smallest Hilbert space \cite{Nielsen:dynamics}. It suffices for tomography to choose the dimension of the ancilla the same as the dimension of the system. After the black box $\mathpzc{E}$ acts on the system, the output state of the total system becomes
\begin{align}
{\varrho}^{(\mathrm{out})}= (\mathpzc{E}\otimes\mathcal{I}) [{\varrho}^{(\mathrm{in})}]=\sum_{\ell} {r}_{\ell} \mathpzc{E}[{A}_{\ell}]\otimes {B}_{\ell},
\label{xx}
\end{align}
where $\mathcal{I}$ denotes the identity operation. In the standard AAPT, measurements on $\varrho^{(\mathrm{out})}$ yield a set of linear equations by solving which one can read $\mathpzc{E}$. We now modify this picture as follows.

\begin{figure}[tp]
\includegraphics[scale=0.3]{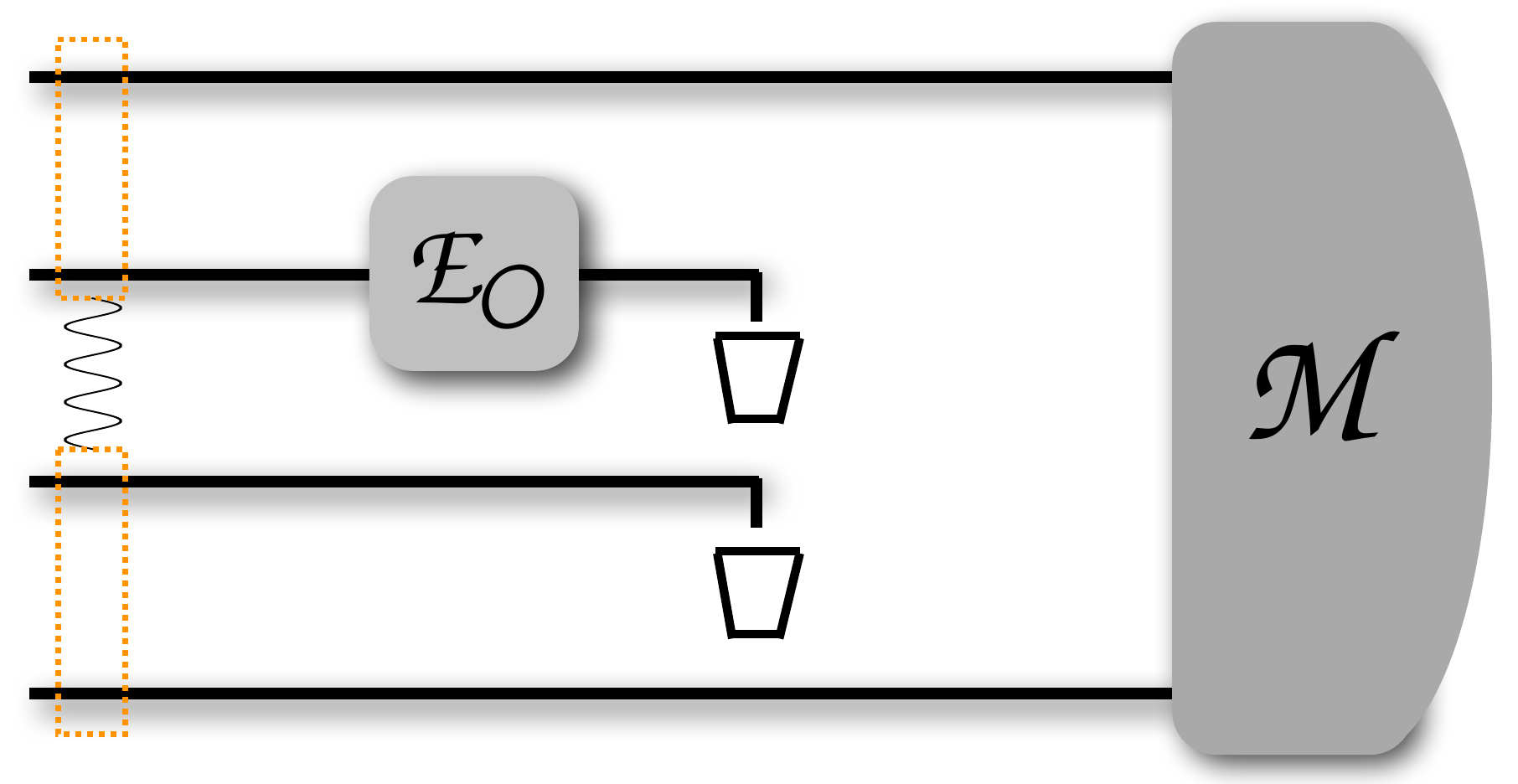} 
\caption{An AAPT scheme with unmeasured particles (discarded). Here $\mathpzc{M}$ denotes measurement on two of the ancillas.
}
\label{aapt}
\end{figure}

The state of the ancilla can be read from $\varrho^{(\mathrm{out})}$ by tracing out over the state of the system,
\begin{align}
{\varrho}_{2}^{(\mathrm{out})}=\sum_{\ell} {r}_{\ell} \mathrm{Tr}\big[\mathpzc{E}[{A}_{\ell}]\big]{B}_{\ell}.
\label{state-out}
\end{align}
This yields
\begin{equation}
\langle {B}^{\dag}_{\ell}\rangle_{\mathrm{out}}={r}_{\ell}\,\mathrm{Tr}\big[\mathpzc{E}[{A}_{\ell}]\big],
\label{linear}
\end{equation}
where $\langle {B}^{\dag}_{\ell}\rangle_{\mathrm{out}} =\mathrm{Tr}[{B}^{\dag}_{\ell}\,\varrho_{2}^{(\mathrm{out})}]$. If ${B}_{\ell}$s are Hermitian operators, this relation gives the result of measuring the \textit{observables} ${B}_{\ell}$ on the final state of the ancilla. If the number of nonzero ${r}_{\ell}$s (i.e., the Schmidt rank) of the initial state is equal to $D^2$ and the number of independent parameters of $\mathpzc{E}$ is not greater than $D^2$, this system of linear equations can be solved \cite{Altepeter,Mohseni}. Thus in this particular case, we see that measurement on the ancilla would be adequate to reproduce the unknown process $\mathpzc{E}$. 

We note, however, that a general quantum channel/operation acting on a system with a $D$-dimensional Hilbert space has $D^4-D^2$ independent parameters \cite{Nielsen}. Hence the above modified picture does not apply to general processes; it applies only when the number of independent parameters in $\mathpzc{E}$ is not greater than $D^2$. An example of such restricted case is depicted in Fig.~\ref{aapt}. Here we have four systems; one on which the (unknown) process $\mathpzc{E}_{\mathpzc{O}}$ acts, and three extra (ancillary) systems, all with the same Hilbert space dimension $d$. For this setting the process $\mathpzc{E}$ of Eq.~(\ref{xx}) is in the form $\mathpzc{E}=\mathpzc{E}_{\mathpzc{O}}\otimes \mathcal{I}$, which has $d^4-d^2$ parameters. That is, here we have $D=d^2$. Note that in this modified AAPT we discard the system (one on which $\mathpzc{E}_{O}$ acts) and one of the ancillas but measure on the remaining two ancillas. In Fig.~\ref{aapt}, $\mathpzc{M}$ denotes measuring the $\{B_{\ell}\}$ observables. 

As a remark, note that if the Schmidt rank of the initial state is $R$ (obviously $\leqslant D^2$) and at the same time the number of unknown parameters of $\mathpzc{E}$ (exactly speaking, $\mathpzc{E}_{O}$) is $\leqslant R$, the modified AAPT method still works.

The above modified AAPT can be generalized further. One can consider that before discarding the undetected parts, a \textit{known} operation $\mathpzc{F}$ is applied on them. Since a quantum operation is linear, the above argument about solving Eq.~(\ref{linear}) still holds but now the equation is modified as 
\begin{equation}
\langle {B}^{\dag}_{\ell}\rangle_{\mathrm{out}}/{r}_{\ell}=\mathrm{Tr}\big[\mathpzc{F}\circ\mathpzc{E}[{A}_{\ell}]\big],
\label{linear-2}
\end{equation}
where ``$\circ$" denotes composition of quantum operations.

\subsection{QI as an AAPT}
\label{subsec:AAPT-UP-2}

\begin{figure}[tp]
\includegraphics[scale=0.3]{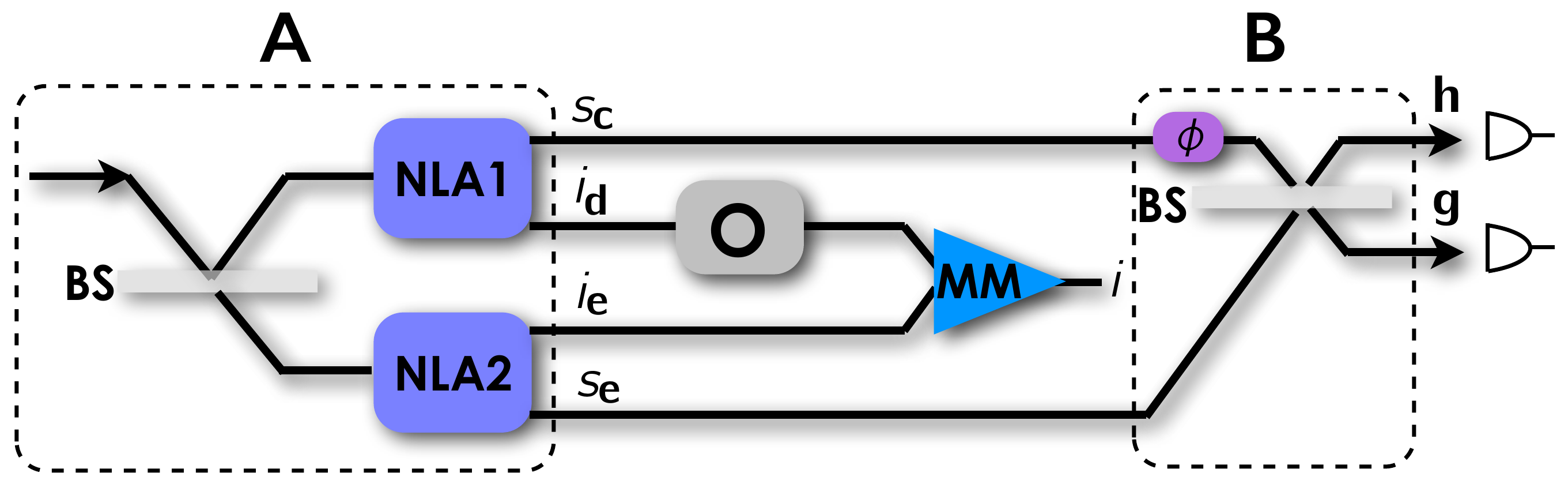}
\caption{(color online). A circuit-like version of the QI process of Fig.~\ref{fig:setup}. Here parts \textbf{\textsf{A}} and \textbf{\textsf{B}} show the state preparation and measurement parts of the corresponding AAPT. Note that Fig.~\ref{fig:setup} does not include the phase shifter $\phi$ (in the measurement part). The role of this extra device is to enable full tomography of the object operation---see Subsec.~\ref{app:complete}.}
\label{qu-im-cir-lik}
\end{figure}

A circuit-like version of the QI setup is sketched in \figref{qu-im-cir-lik}, where the dashed parts \textbf{\textsf{A}} and \textbf{\textsf{B}} represent, respectively, the state preparation and the measurement \cite{Mohseni}. 

For the imaging setup of Fig.~\ref{logic-circuit} (or Fig.~\ref{qu-im-cir-lik}), $\mathpzc{E}_{O}$ ($O$) acts on $i_{1}$ ($i_{\mathbf{d}}$) and has only two independent parameters $T$ and $\gamma$ (App. \ref{obj-map}), and the idlers are undetected (hence $D=2^2$); we only detect the signal particles. Here the initial state is the \textit{entangled} state (\ref{Psi-probe}),
\begin{align}
\varrho^{(\mathrm{in})}_{i_{1}i_{2},s_{1}s_{2}}=|\Phi_{\mathrm{probe}}\rangle\langle\Phi_{\mathrm{probe}}|= \sum_{\ell=1}^{4}r_{\ell} (A_{\ell})_{i_1 i_2} \otimes (B_{\ell})_{s_1 s_2}, \label{Schm}
\end{align}
where $r_{1}=r_{2}=r_{3}=-r_{4}=1/2$ and
\begin{subequations}
\begin{align}
A_{1}&= B_{1}=(1/\sqrt{8})\big(II -ZZ\big),\\
A_{2}&=B_{2} = (1/\sqrt{8}) \big(ZI - IZ\big),\\
A_{3}&=B_{3} = (1/\sqrt{8}) \big( XX+YY\big),\\
A_{4}&=B_{4}=(1/\sqrt{8}) \big(XY - YX\big).
\end{align}
\end{subequations}
Here for brevity we have removed the tensor product symbol (thus, e.g., $XY=X\otimes Y$), and $X$, $Y$, and $Z$ are the Pauli matrices \cite{Nielsen}, and $I$ is the identity operator. Equation (\ref{Schm}) is the Schmidt decomposition of $\varrho^{(\mathrm{in})}$. But note that here we have chosen $r_{4}=-1/2$ in order to make $\{A_{\ell}\}$ and $\{B_{\ell}\}$ (orthonormal) \textit{Hermitian} operators. Although $\varrho^{(\mathrm{in})}$ is not full-rank ($R=4<2^{4}$), it suffices to determine $\mathpzc{E}_{O}$.

After applying the object operation $\mathpzc{E}_{O}$ and the MM operation $\mathpzc{F}_{\mathrm{MM}}$ on $i_{1}i_{2}$, the state of the total $i_{1}i_{2}s_{1}s_{2}$ [Eq.~(\ref{state-out})] becomes
\begin{widetext}
\begin{align}
\Sigma=&\frac{1}{2^{4}}\Big[\mathpzc{F}_{\mathrm{MM}}\circ (\mathpzc{E}_{O}\otimes \mathcal{I})[II -ZZ]\otimes \big(II -ZZ\big) + \mathpzc{F}_{\mathrm{MM}}\circ (\mathpzc{E}_{O}\otimes \mathcal{I})[ZI-IZ]\otimes\big( ZI-IZ\big)\nonumber\\
&+ \mathpzc{F}_{\mathrm{MM}}\circ (\mathpzc{E}_{O}\otimes \mathcal{I})[XX+YY] \otimes \big( XX+YY\big) - \mathpzc{F}_{\mathrm{MM}}\circ (\mathpzc{E}_{O}\otimes \mathcal{I})[XY-YX] \otimes \big( XY-YX\big)\Big] \nonumber\\
=&\frac{\sqrt{2}}{4}\Big( \big[(1+T^2)|\Xi\rangle\langle \Xi|+(1-T^2)|00\rangle 00|\big]\otimes \frac{II-ZZ}{\sqrt{8}} + \big[(1-T^2)|\Xi\rangle\langle \Xi|-(1-T^2)|00\rangle 00|\big]\otimes \frac{ZI-IZ}{\sqrt{8}} \nonumber\\
& -2T\sin\gamma\, |\Xi\rangle\langle \Xi|\otimes \frac{XY-YX}{\sqrt{8}} + 2T\cos\gamma\, |\Xi\rangle\langle \Xi|\otimes \frac{XX+YY}{\sqrt{8}}\Big).
\end{align}

Discarding the $i_{1}i_{2}$ systems (i.e., tracing out over $i_{1}i_{2}$) yields the state of the signal systems ($s_{1}s_{2}$) as
\begin{align}
\Upsilon = \frac{\sqrt{2}}{2} \Big[\frac{II-ZZ}{\sqrt{8}} + T\cos\gamma\,\frac{XX+YY}{\sqrt{8}} - T\sin\gamma\,\frac{XY-YX}{\sqrt{8}} \Big].
\label{eq-Y} 
\end{align}
\end{widetext}
This relation indicates that in order to obtain the object parameters $T$ and $\gamma$, it suffices to measure the observables $XX+YY$ and $XY-YX$. These measurements conclude \textit{full} AAPT of the object. In the quantum circuit of Fig.~\ref{logic-circuit}, the last part (a \textsc{cnot}, a Hadamard, and another \textsc{cnot}) before the detectors, represent BS2 in the experimental setup of Fig.~\ref{setup} or the beam splitter in part \textbf{\textsf{B}} of Fig~\ref{qu-im-cir-lik}. The operators 
\begin{align}
M_{\mathbf{h}}&=|1\rangle_{\mathbf{h}}\langle 1|\otimes |0\rangle_{\mathbf{g}}\langle 0|,\\
M_{\mathbf{g}}&=|0\rangle_{\mathbf{h}}\langle 0|\otimes |1\rangle_{\mathbf{g}}\langle 1|, 
\label{detectors}
\end{align}
indicate the action of the final two detectors---that is, we either detect a photon in path $s_{1}$ ($\mathbf{h}$) or a photon in path $s_{2}$ ($\mathbf{g}$). Hence, the measurement performed on $s_{1}s_{2}$ ($\mathbf{hg}$) can be described by the following two observables:
\begin{align}
\mathpzc{M}_{\mathbf{h}/\mathbf{g}} &= \Big(\textsc{cnot}\,H\otimes I\, \textsc{cnot}\Big) M_{\mathbf{h}/\mathbf{g}} \Big(\textsc{cnot}\, H\otimes I\, \textsc{cnot} \Big)\nonumber\\
&= \frac{\sqrt{2}}{2}\Big[\frac{II-ZZ}{\sqrt{8}} \mp\frac{XX+YY}{\sqrt{8}}\Big].
\label{MEAS}
\end{align}
It is straightforward to see that indeed $\mathpzc{M}_{\mathbf{h}/\mathbf{g}}$ are the following \textit{Bell-state measurements} (App. \ref{BSM}):
\begin{align}
\mathpzc{M}_{\mathbf{h}/\mathbf{g}} &= |\Psi^{\mp}\rangle\langle \Psi^{\mp}|,
\end{align}
where $|\Psi^{\mp}\rangle=(|01\rangle\mp|10\rangle)/\sqrt{2}$.

With these Bell-state measurements, the probability of detecting a photon in $s_{1}$ (path $\mathbf{h}$) is obtained as
\begin{align}
P_{\mathbf{h}}= \mathrm{Tr}[\mathpzc{M}_{\mathbf{h}}\,\Upsilon]=(1-T\cos\gamma)/2,
\end{align}
and similarly for detecting a photon in $s_{2}$ (path $\mathbf{g}$),
\begin{align}
P_{\mathbf{g}}= \mathrm{Tr}[\mathpzc{M}_{\mathbf{g}}\,\Upsilon]=(1+T\cos\gamma)/2,
\end{align}
in agreement with Eq.~(\ref{prob-main-setup}).

Several remarks are in order here. (i) It is seen that $P_{\mathbf{h}} + P_{\mathbf{g}} = 1$, that is we definitely detect one photon either in path $\mathbf{h}$ or in path $\mathbf{g}$. (ii) It is evident that with only $P_{\mathbf{h}}$ or $P_{\mathbf{g}}$ (either or even both), or equivalently the Bell-state measurements $\mathpzc{M}_{\mathbf{h}/\mathbf{g}}$, one cannot obtain the \textit{complete} information of the object (i.e., both $T$ and $\gamma$ at the same time). Hence the QI setup of Fig.~\ref{setup} corresponds to a \textit{partial} AAPT. As we commented after Eq.~(\ref{eq-Y}), one further needs the other (non-Bell-state) measurement $XY-YX$ to fully characterize the object $O$. We discuss later how one can realize this measurement by a simple modification of the original setup of Fig.~\ref{setup}. (iii) It is clear that the initial probe state $|\Phi_{\mathrm{probe}}\rangle$ [Eq.~(\ref{Psi-probe})] is an \textit{entangled} state. We also discuss below how essential this entanglement is for the QI setup to work.

\begin{figure}[tp]
\includegraphics[scale=0.37]{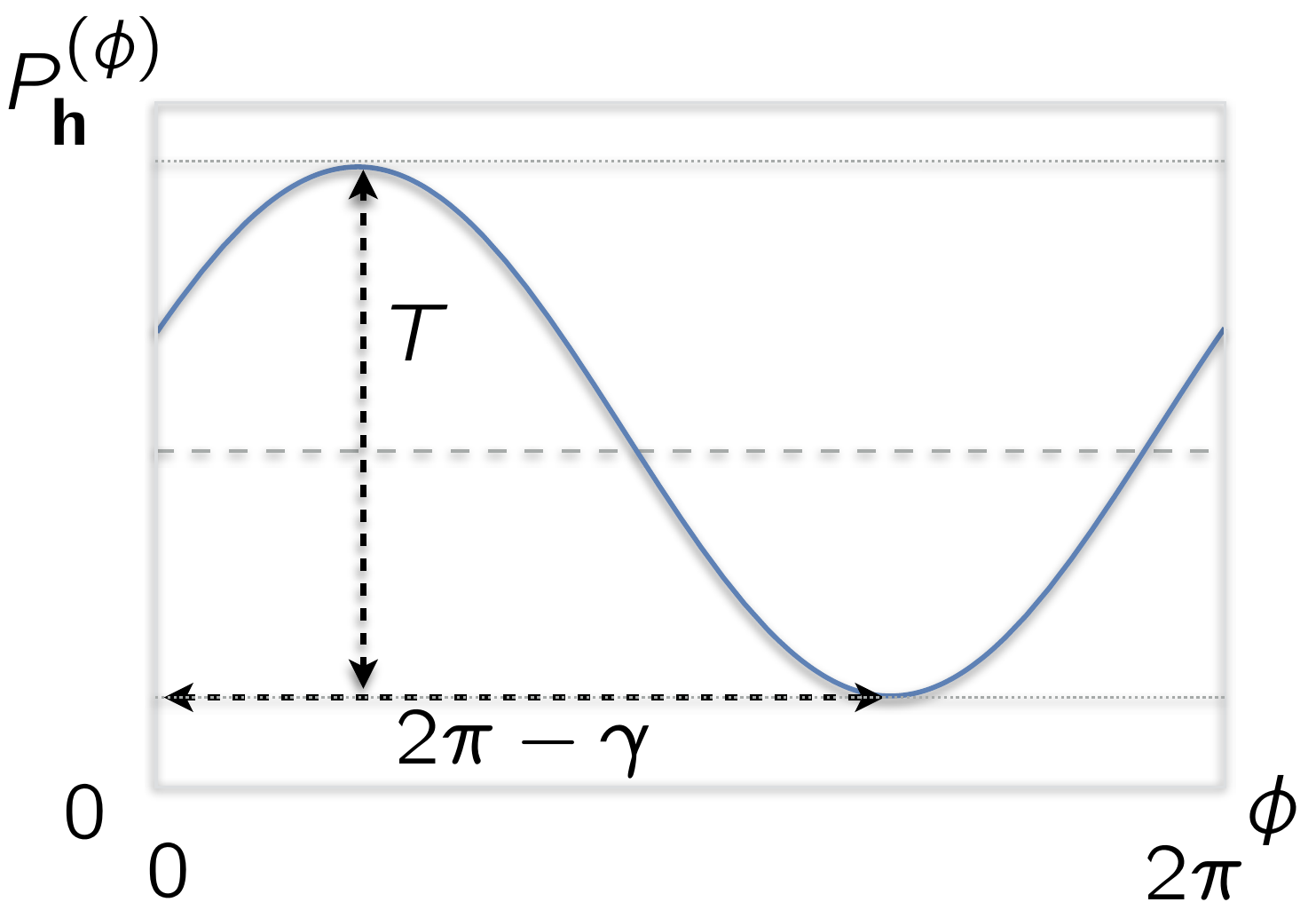}
\caption{Probability of detecting a photon in path $\mathbf{h}$ as a function of the phase parameter $\phi$. This plots shows that how by varying $\phi$ one can read both $T$ and $\gamma$.}
\label{visibility}
\end{figure}

\subsection{Complete tomography of the object}
\label{app:complete}

Here we focus on the dashed part \textbf{\textsf{B}} in \figref{qu-im-cir-lik}, which represents measurement in the QI setup. As we discussed earlier in this section, the final measurement in the QI setup of Fig.~\ref{fig:setup}, which gives the probabilities $P_{\mathbf{h}/\mathbf{g}}$, is tantamount to a partial tomography of the object---in the sense that through this measurement we only obtain the value of $T\cos\gamma$, not both $T$ and $\gamma$ separately. We also argued that a full tomography would require another measurement ($XY-YX$). Here we demonstrate that in order for the full tomography of the object, one needs to slightly modify the existing measurement by adding a controllable phase shifter ($\phi$ in part \textbf{\textsf{B}} of Fig.~\ref{qu-im-cir-lik}). We show that this modification enables measurement of the missing observable. 

One can see that the action of the phase shifter is to change $\mathpzc{M}_{\mathbf{h}/\mathbf{g}}$ to the following observables (App. \ref{app:meas}):
\begin{align}
\mathpzc{M}^{(\phi)}_{\mathbf{h}/\mathbf{g}} =& \frac{\sqrt{2}}{2}\Big[\frac{II-ZZ}{\sqrt{8}} \mp\Big(\cos\phi\,\frac{XX+YY}{\sqrt{8}} \nonumber\\
&-\sin\phi\,\frac{XY-YX}{\sqrt{8}}\Big) \Big].
\end{align}
Hence, in the output of the detectors we obtain the following probabilities (rather than $P_{\mathbf{h}/\mathbf{g}}$):
\begin{align}
P^{(\phi)}_{\mathbf{h}/\mathbf{g}} =\mathrm{Tr}\big[\mathpzc{M}^{(\phi)}_{\mathbf{h}/\mathbf{g}}\,\Upsilon\big]= (1/2)\big[1\mp T\cos(\gamma+\phi)\big].
\end{align}
It is straightforward to see that by varying $\phi$, one can read both $T$ and $\gamma$ through $P^{(\phi)}_{\mathbf{h}}$ (or $P^{(\phi)}_{\mathbf{g}}$) as in Fig.~\ref{visibility}, which is a technique used in the Mach-Zehnder interferometry \cite{Sjq}. 

\subsection{Role of quantum correlation in QI}
\label{app:corr}

It may seem that quantum entanglement in the QI setup of Fig.~\ref{fig:setup} is essential to obtain an image. However, employing the AAPT version of the QI setup allows to see that even separable initial states can be useful in order to extract an image. The fact that entanglement is not essential has already been proved for AAPT \cite{Altepeter}.

As seen from \figref{logic-circuit}, the initial (prepared) state right before the object is $\ket{\Phi_{\mathrm{probe}}}$ [Eq.~(\ref{Psi-probe})], which is an entangled Bell state, which can be represented in a form more familiar for the Bell states as  
\begin{align}
\ket{\Phi_{\mathrm{probe}}}=\frac{1}{\sqrt{2}}  \big(|\mathbf{0}\rangle_{s_1i_1} |\mathbf{1}\rangle_{i_2s_2} + |\mathbf{1}\rangle_{s_1i_1} |\mathbf{0}\rangle_{i_2s_2}\big),
\end{align}
by using the encoding $|\mathbf{0}\rangle_{s_{m} i_{m}}=|0\rangle_{s_m}|0\rangle_{i_m}$ and $|\mathbf{1}\rangle_{s_{m} i_{m}}=|1\rangle_{s_m}|1\rangle_{i_m}$ ($m\in\{1,2\}$). 

We replace this (extended) Bell state with the following (extended) Werner state \cite{Werner}
\begin{equation}
\label{ex-Werner}
{W}_{\mathrm{ext}}=(\xi/4) I\otimes I + (1-\xi)\ketbra{\Phi_{\mathrm{probe}}}{\Phi_{\mathrm{probe}}},
\end{equation}
to see how necessary entanglement is for the QI---see Ref.~\cite{DCQD-corr} for a similar study in QPT and Ref.~\cite{Werner-state:Zhang} for experimental realization of Werner states. The parameter $\xi$ here adjusts the strength of quantum entanglement between paths $s_1i_1$ and $s_2 i_2$ (see Fig.~\ref{aapt}); $W_{\mathrm{ext}}$ becomes entangled for $0\leqslant \xi<2/3$ and separable for $\xi\geqslant 2/3$ \cite{Wootters}.

Running the quantum circuit (\figref{logic-circuit}) by using the initial state (\ref{ex-Werner}) yields the following probabilities:
\begin{equation}
P_{s_{1}/s_{2}}=(1/2)\big[1\mp (1-\xi)T \cos\gamma\big],
\label{Werner-probe}
\end{equation}
which (as expected) for the particular cases of $\xi=0$ and $\xi=1$ give $ (1/2)(1\mp T \cos\gamma )$ and $1/2$, respectively. This relation indicates that, in principle, any probe state with $\xi\neq 1$ can give an image. The quality of the image is quantified by its visibility, defined as $(P_{\max}-P_{\min})/(P_{\max}+P_{\min})$. For $P_{s_{1}/s_{2}}$ the visibility is equal to $(1-\xi)T$, which increases by $T$ and decreases by $\xi$. Thus one can obtain image even with non-entangled initial states. We remark that ``quantum-mimetic" \textit{classical} imaging has already been experimentally demonstrated; see, e.g., Ref.~\cite{Shapiro2}, where an image (of course with relatively smaller visibility) has been obtained by using classical light in an imaging scheme.

\section{Summary}
\label{sec:conc}

We have demonstrated a detailed analogy between the quantum imaging method of Ref.~\cite{Lemos} and ancilla-assisted process tomography. We have suggested a quantum circuit corresponding to the experimental setup, which fully and identically reproduces the results of the imaging scenario and underlies how quantum imaging can be recast in quantum computation/information language. Through the tomography-imaging analogy we have analyzed the utility of non-entangled fields in imaging. In particular, we have argued that it is possible to create an image if the maximally-entangled initial states of the AAPT version of the imaging setup is replaced by a separable Werner state. This implies that the quantum imaging scheme is not totally a result of quantum correlations in the form of entanglement. 

Our work can allow better understand behaviors of different identification or sensing schemes in the form a unified picture.

\textit{Acknowledgments.---}The authors acknowledge useful discussions with G. B. Lemos and A. Zeilinger. A.T.R. was partially supported by Sharif University of Technology's Office of Vice President for Research and Iran Science Elites Federation. 


\begin{widetext}
\appendix

\section{State preparation}
\label{psi_0-psi_prob}

We start our circuit using four qubits in separated $|0\rangle$ modes as the input
\begin{equation}
\ket{\Phi_0}=|0\rangle_{s1}|0\rangle_{i1}|0\rangle_{i2}|0\rangle_{s2},
\end{equation}
where ``$1$" and ``$2$" imply the source of photons (NL1 or NL2). A Hadamard gate followed by a \textsc{cnot} play the role of the first beam splitter (BS1), which is responsible for the uncertainty about the path of the photon (\figref{logic-circuit}). By applying on $\ket{\Phi_0}$, the Hadamard gate generates the following superposition:
\begin{equation}
\ket{\Phi_1}=\big(|0\rangle_{s1}|0\rangle_{i1}|0\rangle_{i2}|0\rangle_{s2}+|0\rangle_{s1} |0\rangle_{i1}|1\rangle_{i2}|0\rangle_{s2}\big)/\sqrt{2}, \nonumber
\end{equation}
which after the \textsc{cnot} gate becomes
\begin{equation}
\ket{\Phi_2}=\big(|0\rangle_{s1}|1\rangle_{i1} |0\rangle_{i2}|0\rangle_{s2}+ |0\rangle_{s1} |0\rangle_{i1}|1\rangle_{i2} |0\rangle_{s2}\big)/\sqrt{2}.
\end{equation}
At the next step, we apply a \textsc{cnot} to simulate the first nonlinear crystal (NL1),
\begin{equation}
\ket{\Phi_3}=\big(|1\rangle_{s1}|1\rangle_{i1}|0\rangle_{i2}|0\rangle_{s2} + |0\rangle_{s1} |0\rangle_{i1} |1\rangle_{i2} |0\rangle_{s2}\big)/\sqrt{2},
\end{equation}
after which another \textsc{cnot} is applied to complete the photon generation process, yielding \eeqref{Psi-probe}.

\section{Object as a quantum channel}
\label{obj-map}

In Ref.~\cite{Lemos}, a semi-transparent object is studied experimentally. We can attribute the following quantum channel (or operation) to the object:
\begin{align}
\mathpzc{E}_O:\begin{cases}|0\rangle_{\mathbf{d}} \rightarrow |0\rangle_{\mathbf{d}}\\ |1\rangle_{\mathbf{d}}\rightarrow Te^{-i\gamma}|1\rangle_{\mathbf{d}} +\sqrt{1-T^2}|1\rangle_{\mathbf{w}}. \end{cases}
\end{align}
Although this shows the correct action of the object, it does not represent an accurate mathematical description. The question is: \textit{what is a mathematically meaningful description of an object which transmits a photon with a factor of $T e^{i\gamma}$, reflects it with $\sqrt{1-T^2}$, and does not give anything when there is no incident photon}?

Here we argue that the object---as a scatterer of a field---can in fact be described as an \textit{amplitude-damping process}, which can be modeled by a BS in the path of a photon \cite{Nielsen,Preskill}. A BS couples the incident field mode $\mathbf{d}$ (represented by the associated annihilation operator ${\mathrm{a}}_{\mathbf{d}}$) to another mode $\mathbf{w}$ (represented by the annihilation operator ${\mathrm{b}}_{\mathbf{w}}$), through the interaction ${H}_{\mathrm{SB}}= i\theta ({\mathrm{a}}_{\mathbf{d}} {\mathrm{b}}^{\dag}_{\mathbf{w}}-{\mathrm{a}}^{\dag}_{\mathbf{d}} {\mathrm{b}}_{\mathbf{w}})$, with the corresponding unitary evolution $U_{\mathrm{BS}}=e^{i {H}_{\mathrm{SB}}}$ \cite{Nielsen,Gerry} (\figref{Ob-as-BS}).

We have
$U_{\mathrm{BS}}~{\mathrm{a}}_{\mathbf{d}}~U_{\mathrm{BS}}^{\dag}=\cos\theta~ e^{-i\omega_{\mathbf{d}}}{\mathrm{a}}_{\mathbf{d}}+\sin\theta ~e^{-i\omega_{\mathbf{w}}}{\mathrm{b}}_{\mathbf{w}}$,
from whence
\begin{align}
U_{\mathrm{BS}}(|0\rangle_{\mathbf{d}} |0\rangle_{\mathbf{w}})=&|0\rangle_{\mathbf{d}} |0\rangle_{\mathbf{w}}, \nonumber \\
U_{\mathrm{BS}}(|1\rangle_{\mathbf{d}} |0\rangle_{\mathbf{w}})=&U_{\mathrm{BS}}~({\mathrm{a}}^{\dag}_{\mathbf{d}}|0\rangle_{\mathbf{d}}|0\rangle_{\mathbf{w}} )  \nonumber  \\
=&\cos\theta~e^{i\omega_{\mathbf{d}}}|1\rangle_{\mathbf{d}} |0\rangle_{\mathbf{w}}+ \sin\theta~e^{i\omega_{\mathbf{w}}} |0\rangle_{\mathbf{d}}|1\rangle_{\mathbf{w}}.
\end{align}
\begin{figure}[bp]
\includegraphics[scale=0.35]{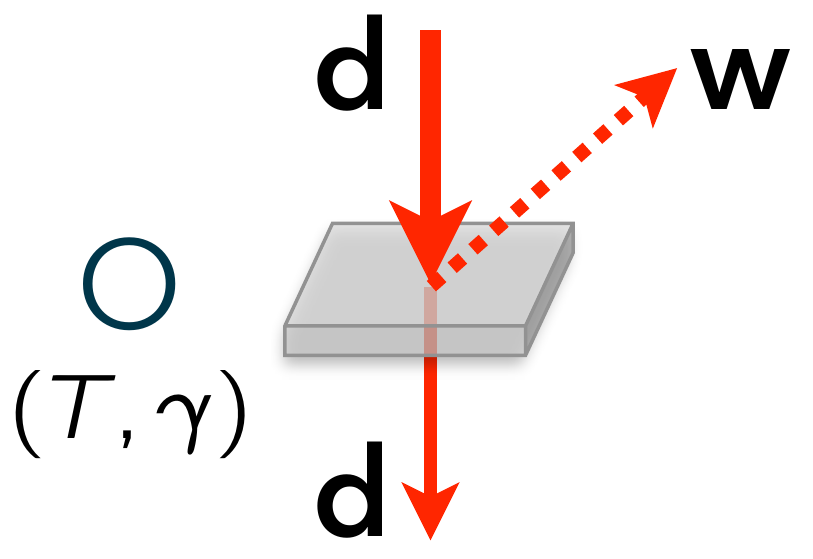}
\caption{(color online). Object---simulated as a beam splitter---transmits the incident beam with probability $\cos^2\theta$ and reflects it otherwise.}
\label{Ob-as-BS}
\end{figure}
By tracing out over mode $\mathbf{w}$ we obtain the following Kraus operators for the operation of the object:
\begin{equation}
\label{kraus}
K_0=
\left(\begin{array}{cc}
1 & 0  \\
0 & \sqrt{\eta}
\end{array}\right) ~\text{and}~~K_1=
\left(\begin{array}{cc}
0 & \sqrt{1-\eta} \\
0 & 0
\end{array}\right),
\end{equation}
where $|\eta |:=\cos^2\theta$, and we have the trace-preserving condition $K_0^{\dag}K_0+K_1^{\dag}K_1=I$ \cite{Nielsen,Preskill}. The object may also change the phase of the incident field; thus to be more general we assume $\sqrt{\eta}=Te^{-i\gamma}$.

The action of the Kraus representation $\mathpzc{E}_O[{\varrho}]=\sum_{\beta} K_{\beta}{\varrho}K_{\beta}^{\dag}$ on the input state
\begin{equation}
{\varrho}=
\left(\begin{array}{cc}
{\varrho}_{00} & {\varrho}_{01}  \\
{\varrho}_{10} & {\varrho}_{11}
\end{array}\right)
\end{equation}
gives the final state as
\begin{equation}
\label{map-kraus}
\mathpzc{E}_O[{\varrho}]=
\left(\begin{array}{cc}
{\varrho}_{00}+(1-T^2){\varrho}_{11} & Te^{-i\gamma}{\varrho}_{01}  \\
Te^{i\gamma}{\varrho}_{10} &T^2 {\varrho}_{11}
\end{array}\right).
\end{equation}
One can also find a more explicit representation for this process by expanding $K_{0}$ and $K_{1}$ in terms of the orthonormal basis operators $\{\sigma_{0}\equiv I,\sigma_{1}\equiv \sigma_{x},\sigma_{2}\equiv \sigma_{y},\sigma_{3}\equiv \sigma_{z}\}/\sqrt{2}$, where $\sigma_{\alpha}$s are the Pauli matrices \cite{nb}. Specifically, defining the coefficients $a_{\beta\alpha}$ through $K_{\beta}=\sum_{\alpha=0}^{3}a_{\beta\alpha}\sigma_{\alpha}/\sqrt{2}$ yields $\mathpzc{E}_{O}[\varrho]=(1/2)\sum_{\alpha\beta}(\chi_{O})_{\alpha\beta} \,\sigma_{\alpha}\varrho\sigma_{\beta}$, where $(\chi_{O})_{\alpha\beta} = \sum_{l=0}^{3}a_{l\alpha}a^{*}_{l\beta}$ can be considered a matrix representation of $\mathpzc{E}_{O}$ in the orthonormal operator basis $(1/\sqrt{2})\{\sigma_{\alpha}\}_{\alpha=0}^{3}$. Thus we obtain
\begin{equation}
\chi_{O}=\left(\begin{array}{cccc} |1+\sqrt{\eta}|^2 & 0 & 0 & (1+\sqrt{\eta})(1-\sqrt{\eta}^{*}) \\ 0 & |\sqrt{1-\eta}|^{2} & i|\sqrt{1-\eta}|^{2} & 0\\ 0 & i|\sqrt{1-\eta}|^{2} & |\sqrt{1-\eta}|^{2} & 0 \\ (1+\sqrt{\eta}^{*})(1-\sqrt{\eta}) & 0 & 0 & |1-\sqrt{\eta}|^{2}  \end{array}\right).
\end{equation}
Note that although $\chi_{O}$ is a $4\times 4$ positive-definite matrix with $(1/2)\sum_{\alpha\beta}(\chi_{O})_{\alpha\beta}\sigma_{\beta}\sigma_{\alpha}=I$ (the trace-preserving condition), it only has $2$ independent real parameters $T$ and $\gamma$.

\section{Applying the object and MM operations}
\label{object-app}

Applying $\mathpzc{E}_{\mathpzc{O}}$ [\eeqref{map}] on the following state:
\begin{align}
{\varrho}_2\equiv& |\Psi_{\mathrm{probe}} \rangle\langle \Psi_{\mathrm{probe}} | \nonumber\\
=& \frac{1}{2}\big(\ketbra{1}{1} \otimes \ketbra{1}{1} \otimes \ketbra{0}{0} \otimes \ketbra{0}{0}  + \ketbra{1}{0} \otimes \ketbra{1}{0} \otimes \ketbra{0}{1} \otimes \ketbra{0}{1} \nonumber \\
&+ \ketbra{0}{1} \otimes \ketbra{0}{1} \otimes \ketbra{1}{0} \otimes \ketbra{1}{0}  + \ketbra{0}{0} \otimes \ketbra{0}{0} \otimes \ketbra{1}{1} \otimes \ketbra{1}{1} \big)
\end{align}
yields
\begin{align}
\label{a}
{\varrho}_3\equiv & (\mathcal{I}\otimes\mathpzc{E}_{O}\otimes \mathcal{I} \otimes \mathcal{I})[{\varrho}_2] \nonumber\\
=&  \frac{1}{2}\big(\ketbra{1}{1} \otimes \mathpzc{E}_O[\ketbra{1}{1}] \otimes \ketbra{0}{0} \otimes \ketbra{0}{0}  + \ketbra{1}{0} \otimes \mathpzc{E}_O[\ketbra{1}{0}] \otimes \ketbra{0}{1} \otimes \ketbra{0}{1} \nonumber \\
&+ \ketbra{0}{1} \otimes \mathpzc{E}_O[\ketbra{0}{1}] \otimes \ketbra{1}{0} \otimes \ketbra{1}{0}  + \ketbra{0}{0} \otimes \mathpzc{E}_O(\ketbra{0}{0}) \otimes \ketbra{1}{1} \otimes \ketbra{1}{1} \big) \nonumber \\
=& \frac{1}{2} \big(T^2\ketbra{1}{1} \otimes \ketbra{1}{1} \otimes \ketbra{0}{0} \otimes \ketbra{0}{0} + (1-T^2) \ketbra{1}{1} \otimes \ketbra{0}{0} \otimes \ketbra{0}{0} \otimes \ketbra{0}{0} \nonumber \\
&+ Te^{i\gamma}\ketbra{1}{0} \otimes \ketbra{1}{0} \otimes \ketbra{0}{1} \otimes \ketbra{0}{1} + Te^{-i\gamma} \ketbra{0}{1} \otimes \ketbra{0}{1} \otimes \ketbra{1}{0} \otimes \ketbra{1}{0}   \nonumber \\
&+ \ketbra{0}{0} \otimes \ketbra{0}{0} \otimes \ketbra{1}{1} \otimes \ketbra{1}{1} \big).
\end{align}

After applying MM [Eq.~(\ref{MM-def})], we obtain
\begin{align}
\label{before-M0}
{\Sigma}=& \frac{1}{2}\big[ T^2 \ketbra{1}{1}\otimes |\Xi\rangle\langle \Xi|  \otimes \ketbra{0}{0} +(1- T^2) \ketbra{1}{1}\otimes |00\rangle\langle 00| \otimes \ketbra{0}{0} + Te^{i\gamma}\ketbra{1}{0}\otimes |\Xi\rangle\langle \Xi| \otimes \ketbra{0}{1}   \nonumber \\
&+ Te^{-i\gamma} \ketbra{0}{1}\otimes |\Xi\rangle\langle \Xi| \otimes \ketbra{1}{0} + \ketbra{0}{0}\otimes |\Xi\rangle\langle \Xi| \otimes \ketbra{1}{1}\big].
\end{align}

Now if we discard the idler photons, which means tracing out over $i_{1} i_{2}$, the state of the signal photons ($s_{1} s_{2}$) reduces to 
\begin{align}
\Upsilon=\frac{1}{2}\big[|1\rangle\langle1|\otimes |0\rangle\langle 0| + T e^{i\gamma} |1\rangle\langle 0| \otimes |0\rangle \langle 1| + T e^{-i\gamma}|0\rangle \langle 1|\otimes |1\rangle \langle 0| + |0\rangle\langle 0|\otimes |1\rangle\langle 1|  \big].
\end{align}
Note that this is exactly the state $\varrho_{6}$ [Eq.~(\ref{bef-BS2})] of the original experimental setup.

\section{Bell-state measurements}
\label{BSM}

We recall the definition of the Bell states for the case of two qubits \cite{Nielsen},
\begin{align}
|\Phi^{\pm}\rangle&=\frac{1}{\sqrt{2}}(|00\rangle \pm|11\rangle),\\
|\Psi^{\pm}\rangle&=\frac{1}{\sqrt{2}}(|01\rangle\pm|10\rangle).
\end{align}
It is straightforward to verify the following Schmidt decompositions for the Bell-state measurements:
\begin{align}
|\Phi^{+}\rangle\langle \Phi^{+}| =(II+XX-YY+ZZ)/4,\\
|\Phi^{-}\rangle\langle \Phi^{-}| =(II-XX+YY+ZZ)/4,\\
|\Psi^{+}\rangle\langle \Psi^{+}| =(II+XX+YY-ZZ)/4,\\
|\Psi^{-}\rangle\langle \Psi^{-}| =(II-XX-YY-ZZ)/4.
\end{align}

\section{Measurements for the complete object tomography}
\label{app:meas}

A phase shifter $Z_{\phi}=\left(\begin{smallmatrix}1 & 0\\ 0 & e^{i\phi} \end{smallmatrix} \right)$ has the following action on the $X$ and $Y$ Pauli matrices:
\begin{align}
Z_{\phi}XZ_{\phi}^{\dag} & = \cos \phi\, X + \sin \phi\, Y, \label{Z-1}\\
Z_{\phi}YZ_{\phi}^{\dag} & = -\sin \phi\, X + \cos \phi\, Y. \label{Z-2}
\end{align}
As a result, one can see that adding $Z_{\phi}$ before the BS and detectors in part \textbf{\textsf{B}} of Fig.~\ref{qu-im-cir-lik} changes $\mathpzc{M}_{\mathbf{h}/\mathbf{g}}$ to
\begin{align}
\mathpzc{M}^{(\phi)}_{\mathbf{h}/\mathbf{g}} &= \Big(Z_{\phi}\otimes I\,\textsc{cnot}\,H\otimes I\, \textsc{cnot}\Big) M_{\mathbf{h}/\mathbf{g}} \Big(\textsc{cnot}\, H\otimes I\, \textsc{cnot}\, Z^{\dag}_{\phi}\otimes I\Big)\nonumber\\
&\overset{(\ref{MEAS})}{=} \frac{1}{4}(Z_{\phi}\otimes I)\big[(II-ZZ)\mp(XX+YY) \big] (Z^{\dag}_{\phi} \otimes I)\nonumber\\
&\overset{(\mathrm{\ref{Z-1}}),(\mathrm{\ref{Z-2}})}{=} \frac{\sqrt{2}}{2}\Big[ \frac{II-ZZ}{\sqrt{8}} \mp \Big(\cos\phi\,\frac{XX+YY}{\sqrt{8}}-\sin\phi\,\frac{XY-YX}{\sqrt{8}}\Big)\Big].
\end{align}
Hence we see that this phase shifter allows the missing measurement $XY-YX$ appear.

\twocolumngrid
\end{widetext}


\begin{thebibliography}{99}

\bibitem{Klyshko-1}D. N. Klyshko,  Usp. Fiz. Nauk. \textbf{154}, 133 (1988); 
%
Sov. Phys. Usp. \textbf{31}, 74 (1988); 
%
Phys. Lett. A \textbf{132}, 299 (1988).

\bibitem{Pittman} T. B. Pittman, Y. H. Shih, D. V. Strekalov, A. V. Sergienko, Phys. Rev. A \textbf{52}, R3429 (1995).

\bibitem{Gatti-1} A. Gatti, E. Brambilla, M. Bache, and L. A. Lugiato, Phys. Rev. A \textbf{70}, 013802 (2004).

\bibitem{Shih} Y. Shih, IEEE J. Sel. Top. Quantum Electron. \textbf{13}, 1016 (2007).

\bibitem{Shapiro} J. H. Shapiro, Phys. Rev. A \textbf{78}, 061802(R) (2008).

\bibitem{Aspden}R. S. Aspden, D. S. Tasca, R. W. Boyd, and M. J. Padgett, New J. Phys. \textbf{15}, 073032 (2013).

\bibitem{Bennink} R. S. Bennink, S. J. Bentley, and R. W. Boyd, Phys. Rev. Lett. \textbf{89}, 113601 (2002).

\bibitem{Gatti-2} A. Gatti, E. Brambilla, M. Bache, and L. A. Lugiato, Phys. Rev. Lett. \textbf{93}, 093602 (2004).

\bibitem{Valencia} A. Valencia, G. Scarcelli, M. D'Angelo, and Y. Shih, Phys. Rev. Lett. \textbf{94}, 063601 (2005).

\bibitem{Ferri}F. Ferri, D. Magatti, A. Gatti, M. Bache, E. Brambilla, and L. A. Lugiato, Phys. Rev. Lett. \textbf{94}, 183602 (2005).

\bibitem{which-way} K. Banaszek, P. Horodecki, M. Karpi\'{n}ski, and C. Radzewicz, Nature Commun. \textbf{4}, 2594 (2013).

\bibitem{Lemos}G. B. Lemos, V. Borish, G. D. Cole, S. Ramelow, R. Lapkiewicz, and A. Zeilinger, Nature \textbf{512}, 409 (2014).

\bibitem{Lahiri}M. Lahiri, R. Lapkiewicz, G. B. Lemos, and A. Zeilinger, Phys. Rev. A \textbf{92}, 013832 (2015).

\bibitem{Nielsen} M. A. Nielsen and I. L. Chuang, \emph{Quantum Computation and Quantum Information} (Cambridge University Press, Cambridge, 2010).

\bibitem{Chuang}I. L. Chuang and M. A. Nielsen, J. Mod. Opt. \textbf{44}, 2455 (1997).

\bibitem{Poyatos}J. F. Poyatos, J. I. Cirac, and P. Zoller, Phys. Rev. Lett. \textbf{78}, 390 (1997).

\bibitem{Leung} D. W. Leung, J. Math. Phys. \textbf{44}, 528 (2003).

\bibitem{D'Ariano-AAPT-1} G. M. D'Ariano and P. Lo Presti, Phys. Rev. Lett. \textbf{86}, 4195 (2001).

\bibitem{Altepeter}J. B. Altepeter, D. Branning, E. Jeffrey, T. C. Wei, P. G. Kwiat, R. T. Thew, J. L. O'Brien, M. A. Nielsen, and A. G. White, Phys. Rev. Lett. \textbf{90}, 193601 (2003).

\bibitem{D'Ariano-AAPT-2} G. M. D'Ariano and P. Lo Presti, Phys. Rev. Lett. \textbf{91}, 047902 (2003).

\bibitem{Mohseni-DCQD} M. Mohseni and D. A. Lidar, Phys. Rev. Lett. \textbf{97}, 170501 (2006).

\bibitem{DCQD-corr} M. Mohseni, A. T. Rezakhani, J. T. Barreiro, P. G. Kwiat, and A. Aspuru-Guzik, Phys. Rev. A \textbf{81}, 032102 (2010).

\bibitem{Wang-DCQD} Z. W. Wang, Y. S. Zhang, Y. F. Huang, X. F. Ren, and G. C. Guo, Phys. Rev. A \textbf{75}, 044304 (2007).

\bibitem{Mohseni} M. Mohseni, A. T. Rezakhani, and D. A. Lidar, Phys. Rev. A \textbf{77}, 032322 (2008).

\bibitem{Miquel-etal-Nature} C. Miquel, J. P. Paz, M. Saraceno, E. Knill, R. Laflamme, and C. Negrevergne, Nature \textbf{418}, 59 (2002).

\bibitem{bib:NL} For our analysis in this paper, this simplified working of NL suffices. A comprehensive analysis of NL has already been investigated, e.g., in T. J. Herzog, J. G. Rarity, H. Weinfurter, and A. Zeilinger, Phys. Rev. Lett. \textbf{72}, 629 (1994)
 and S. P. Walborn, C. H. Monken, S. P\'{a}dua, and P. H. Souto Ribeiro, Phys. Rep. \textbf{495}, 87 (2010).

\bibitem{Nielsen:dynamics} M. A. Nielsen, C. M. Dawson, J. L. Dodd, A. Gilchrist, D. Mortimer, T. J. Osborne, M. J. Bremner, A. W. Harrow, and A. Hines, Phys. Rev. A \textbf{67}, 052301 (2003).

\bibitem{Sjq} E. Sj\"{o}qvist, A. K. Pati, A. Ekert, J. S. Anandan, M. Ericsson, D. K. L. Oi, and V. Vedral, Phys. Rev. Lett. \textbf{85}, 2845  (2000).

\bibitem{Werner} R. F. Werner, Phys. Rev. A \textbf{40}, 4277 (1989).

\bibitem{Werner-state:Zhang} Y.-S. Zhang, Y.-F. Huang, C.-F. Li, and G.-C. Guo, Phys. Rev. A \textbf{66}, 062315 (2002).

\bibitem{Wootters} W. K. Wootters, Phys. Rev. Lett. \textbf{80}, 2245 (1998).

\bibitem{Shapiro2} J. H. Shapiro, D. Venkatraman, and F. N. C. Wong, Sci. Rep. \textbf{5}, 10329 (2015).

\bibitem{Preskill} J. Preskill, \emph{Quantum Computation: Lecture Notes for Physics 219} (California Institute of Technology, 1998), \url{http://www.theory.caltech.edu/~preskill/ph219/}.

\bibitem{Gerry} C. Gerry and P. Knight, \emph{Introductory Quantum Optics} (Cambridge University Press, Cambridge, 2005).

\bibitem{nb} Note that within the main text, for brevity, we have used a different notation for the Pauli matrices $X\equiv \sigma_{x}$, $Y\equiv \sigma_{y}$, and $Z\equiv \sigma_{z}$.

\end{thebibliography}
\end{document}